\documentclass{osa-article}
\journal{oe}
\articletype{Research Article}
\usepackage{amssymb}
\usepackage{amsmath}
\usepackage{amsfonts}
\usepackage{bm}%

\newcommand*{\rttensor}[1]{\overline{\overline{#1}}}
\newcommand{\var}[1]{{\operatorname{\mathit{#1}}}}
 \DeclareMathOperator\arctanh{arctanh}

\begin{document}
\title{Waveguide Modes in Weyl Semimetals with Tilted Dirac Cones}
\author{Klaus Halterman\authormark{1,*} and Mohammad Alidoust\authormark{2,$\dagger$}}
\address{\authormark{1}Michelson Lab, Physics Division, Naval Air Warfare Center, China Lake, California 93555\\
\authormark{2}Department of Physics, K.N. Toosi University of Technology, Tehran 15875-4416, Iran\\
\email{\authormark{*} klaus.halterman@navy.mil}
\authormark{$\dagger$}phymalidoust@gmail.com}

\begin{abstract}
We theoretically study unattenuated electromagnetic guided wave modes in 
centrosymmetric Weyl semimetal layered systems. By solving Maxwell's equations for the
electromagnetic fields and using the
appropriate boundary conditions, we derive dispersion relations for propagating modes in a finite-sized Weyl semimetal. 
Our findings reveal that for ultrathin structures, and proper Weyl cones tilts, 
extremely localized guided waves can propagate along the semimetal interface
over a certain range of frequencies.
This follows from the anisotropic nature of the semimetal where
the  
diagonal  components of the permittivity can exhibit a
tunable epsilon-near-zero response. From the dispersion diagrams,
we determine experimentally accessible regimes that lead to high
energy-density 
confinement
 in the Weyl semimetal layer. Furthermore, we show that the net system power
can vanish all together, 
depending on the Weyl cone tilt and frequency of the electromagnetic wave.
These effects are seen in
 the  energy transport velocity, which 
 demonstrates a substantial 
 slowdown in the propagation of electromagnetic energy near
 critical points of the dispersion diagrams. Our results can provide 
 guidelines in designing 
 Weyl semimetal waveguides that can offer efficient control 
 in the velocity and direction of energy flow.
\end{abstract}



\section{Introduction}\label{intro}
With the recent discoveries 
of a number of Weyl\cite{weyl}
semimetal (WS) compounds\cite{typ2_2,Savrasov,hsn3,dai,yan,hau,typ2_8,typ2_7,typ2_5,typ2_6,nally,burkov,wyl1,wyl6,wyl7,rev1,rev2}
and
their intriguing optical properties 
\cite{qchen,tama,kot,kats,Kotov1,karg,collect,muk1,muk2,zyuzin,tim,noh,qiu1,burkov,opt1,ashby,ashby2,Sushkov1,wang}, 
there has been
a surge in research that investigates
their potential roles in future
optical systems. When studying the electromagnetic (EM) properties of  WS systems,
it is important to take into consideration their inherent 
anisotropic nature.
Depending on the  
polarization of the EM 
wave relative to the crystalline axes,
a WS can be tuned 
to exhibit a metallic or insulating type of EM response,
as  
described by
its  permittivity  tensor $\rttensor{\epsilon}$. 
This is reflected in the real parts of the appropriate components of 
$\rttensor{\epsilon}$ changing sign, or vanishing altogether, as
occurs 
in epsilon-near-zero (ENZ) materials \cite{cali,leaky,enghetta,alu,alu2,liu,funnel,feng,smol,matt,liberal,colors,HAZ}.
Various peculiar effects can arise in
materials 
exhibiting  an ENZ response \cite{liberal},
including 
light tunneling, vanishing group velocity, 
and
perfect absorption \cite{feng}.
By exploiting  
the anisotropic tunable EM response  of 
Weyl semimetals, 
new classes of optical  devices, such as waveguides,
can be
designed that afford dissipationless,
highly localized energy transport
 with minimal 
attenuation and signal distortion.

For anisotropic waveguides, the  
relative signs of the components of $\rttensor{\epsilon}$
can  also be crucial for the optimization of energy transport.
For example,  a uniaxial anisotropic system that has permittivities of opposite sign can have
an isofrequency surface in the form of an open hyperboloid. Such hyperbolic metamaterials 
have been discussed in the context 
of waveguides\cite{podd}
to generate forward and backward 
surface modes \cite{sun},
and  to enhance  EM fields \cite{he}.
Thus, the anisotropic EM response of a WS
must be accounted for when discussing
its role as a waveguide
in the ENZ regime. Indeed, the  
 sign differences that can arise 
 between the diagonal permittivity components, 
can afford
greater control and increased
subwavelength EM field confinement.

In addition to the dielectric response of Weyl semimetals, 
the tilting of the Weyl cones
can induce observable effects in
the conductivity and polarization \cite{collect}.
By reversing the conical tilt direction, 
the right-hand and left-hand responses of the
 Weyl semimetal become reversed, 
 which can affect the absorption
 of circularly polarized light \cite{muk1}.
 It was found that the chirality or sign of the tilt parameter  
 in the absence of time-reversal and inversion 
 symmetries can change the sign of the 
  Weyl contributions to 
 the absorptive Hall conductivity \cite{muk2}.
 For  
 Weyl semimetal structures much thinner than the incident wavelength,  
 EM radiation can be perfectly absorbed  for 
nearly any incident angle, depending on the tilt of the Weyl cones 
and chemical potential \cite{HAZ}.
Moreover, the broad tunability of the chemical potential in a WS 
makes it a promising material for
photonics and 
plasmonics  
applications\cite{qi1,zhou1,Kharzeev1,Hofmann1,zyuzin,Sushkov1,valla,wang,hau,tim}. 
The chiral anomaly in a WS
can alter surface plasmons and
the EM response \cite{qi1,zhou1,Kharzeev1,Hofmann1,zyuzin,wang,Sushkov1}.
Other interesting optical effects can also arise in Weyl semimetals, 
including the Imbert-Fedorov effect, which can result 
in detectable valley-dependent variations in the velocites of Weyl fermions \cite{xie}.
It has been shown theoretically \cite{Sushkov1}  that
measurements of the optical conductivity 
and the  temperature dependence of the 
free carrier response in pyrochlore
Eu$_2$Ir$_2$O$_7$ is consistent with the WS phase, and
that  the interband optical conductivity
reduces to zero in a continuous fashion
at low frequencies as predicted for a WS.
The surface
magnetoplasmons
of a Weyl semimetal can turn to  low-loss localized guided
modes when two crystals  of the WSs
with different magnetization orientations are
connected \cite{zyuzin}.

In this paper, we investigate 
guided wave modes that
can arise in an anisotropic  Weyl semimetal waveguide
structure.
We consider a finite-sized Weyl semimetal layer on top of a 
perfectly conducting substrate. Our study includes both analytic and numerical results that 
 determine the EM fields and characteristics of the energy flow
 in WS waveguide systems. 
 We consider  broad ranges of conical tilting,
 and EM wave frequencies.
We show that by appropriately tuning the 
frequency and conical tilt, 
both of 
the 
diagonal 
components of $\rttensor{\epsilon}$ can achieve a
lossless  ENZ response. 
From the  
obtained dispersion relations, 
which 
reveal the permitted EM solutions, 
we construct the EM fields 
and identify
regions of  high
energy density 
confinement
 in the WS layer. 
 We demonstrate 
 controllable net 
 system power 
 that can be made to 
 vanish at critical points of the dispersion curves,
where the propagation of electromagnetic energy 
 comes to a standstill. 
 Our findings offer systematic ways to
 design waveguide 
structures  that  
can continuously 
reduce the energy velocity from a fraction of the speed of light 
to  effectively zero.
%
 
 The paper is organized as follows:
 In Sec.~\ref{subsec1}, we summarize the dielectric response of the WS by
 presenting each component of the permittivity tensor, and
 give details of the approximations made. 
 Next, in Sec.~\ref{subsec2}, we solve Maxwell's equations and obtain 
 the dispersion relations for the corresponding guided EM wave modes. 
 We also define the averaged energy density, Poynting vector, and  velocity of energy flow. 
 In Sec.~\ref{subsec3}, we examine the obtained analytical expressions 
 numerically and discuss our main results. 
 Finally
 in Sec.~\ref{conclusion}, we give concluding remarks.

\section{Model and results}\label{methods}
In this section, we outline
the salient features of
the EM response of
a WS by providing analytical expressions
for the components of the permittivity tensor, as well as
the  numerical approach used. 
For clarity, we have considered the zero temperature case, and neglect the influence of 
impurities and disorder. At finite temperatures, or the presence of impurities and disorder, 
the  mode dispersion diagram would acquire additional radiative mode characteristics. For high enough 
temperatures, optical absorption and propagating modes would be dominated less by coherent interference 
effects and more by dissipative ones, adversely impacting the findings presented here.
The
Weyl semimetal  is modeled as a system 
with broken time-reversal symmetry and two Weyl nodes. 
This model can be achieved through 
the stacking of multiple thin films involving a topological 
insulator and ferromagnet blocks \cite{burkov}.
The Hamiltonian describing the low energy physics around the two Weyl nodes,
defined by ``$s=\pm$'', is given by: 
\begin{equation}
H_{s}(\mathbf{p}) = v_F[\beta_s (p_z-sQ) +s {\bm \sigma}(\textbf{p}-sQ\textbf{e}_z)].
\end{equation}
Here  $\textbf{e}_z$ is
the unit vector along the $z$ direction, 
and we take the Fermi velocity $v_F$ to be positive. 
The separation between two Weyl points in the $z$ direction in momentum space is defined by $2|Q|$,
where the sign of $Q$ depends on the sign of the magnetization. 
The tilt of the Weyl cones are described by the parameters  $\beta_{\pm}$,
and for centrosymmetric materials with broken time reversal symmetry, 
we apply the condition  $\beta=\beta_+ = - \beta_-$.

\subsection{Permittivity tensor}\label{subsec1}

The permittivity tensor $\rttensor{\epsilon}$ for the WS takes the
following gyrotropic form:\cite{HAZ}
\begin{eqnarray}\label{rttepsilon}
\rttensor{\epsilon}(\omega)
=\left( \begin{array}{ccc}
\epsilon_{\parallel}(\omega) &i\gamma (\omega)& 0 \\
-i\gamma (\omega)& \epsilon_{\parallel}(\omega)  & 0 \\
0 &0 &\epsilon_{zz}(\omega)  \end{array}\right),
\end{eqnarray}
where
$\epsilon_\parallel(\omega)$ represents  
the equal components of  ${\rttensor{\epsilon}}(\omega)$ parallel to the interfaces,
i.e., $\epsilon_{xx}(\omega)=\epsilon_{yy}(\omega)=\epsilon_\parallel(\omega)$.
Here the components are normalized by $\epsilon_0$, the permittivity of free space. The $\epsilon_{\parallel}(\omega)$ components  can
be written analytically as \cite{HAZ},
\begin{align}
\label{dispexx}
\epsilon_{\parallel}(\omega)=1+
\frac{\alpha}{3\pi}
\left[\text{ln} \left|\dfrac{4\Gamma^2}{4 \mu^2-\omega^2}\right| 
-\dfrac{4 \mu^2}{\omega^2}+i\pi\Theta(\omega-2 \mu)\right],
\end{align}
where $\mu$ is the chemical potential, with  
$\mu\geq 0$, $\alpha=e^2/(4\pi\epsilon_0 \hbar v_F)$, and
$\Theta(X)$ represents the usual step function.
Generally, the energy cut-off $\Gamma$ is a function of the tilt parameter. 
Nevertheless, in our calculations, 
we choose a large enough cut-off and neglect the contribution of $\beta_\pm$ to $\Gamma$.  

It is evident that $\epsilon_{\parallel}(\omega)$ is independent of the tilting
parameters $\beta_{\pm}$. 
For the situation when the Weyl cones are not tilted ($\beta_\pm=0$), 
 we have $\epsilon_{zz}(\omega)=\epsilon_{\parallel}(\omega)$,
and the off-diagonal frequency dependent 
component $\gamma(\omega)$ reduces to,
$\gamma(\omega) =2v_F \alpha Q(\pi\omega)^{-1}$. The imaginary term in Eq.~(\ref{dispexx})
describes the interband contribution to the optical
conductivity, which exists only when
 the frequency $\omega$ of the EM wave
satisfies  $ \omega>2\mu$. 
For the guided waves of interest,
we are interested in the situation where there is no dissipation
in the medium.
As Eq.~(\ref{dispexx}) shows, the range
of frequencies in which 
$\epsilon_{\parallel}$ is real and positive corresponds to
\begin{align}
2\mu\sqrt{\frac{\alpha}{3\pi+{2\alpha}\ln{\big|{\Gamma}/{\mu}\big|}}}<  \omega < 2\mu.
\end{align}

Similarly, the permittivity component $\epsilon_{zz}$ depends  on frequency and chemical potential,
however it also depends on the tilt $\beta_\pm$. 
The  $\epsilon_{zz}$ component involves a complicated integral, and its derivation is
presented elsewhere \cite{HAZ}.
For
 frequencies satisfying  
 \begin{align}
 \omega<\frac{2\mu}{(1+|\beta|)}, 
 \end{align}
 there is no dissipation (for all diagonal components), 
 and $\epsilon_{zz}$ 
 can be
  written analytically as \cite{HAZ},
\label{bigezz}
\begin{align} 
\label{bigezz'}
\epsilon_{zz}=1+\frac{\alpha\mu^2}{\pi\omega^2}&\sum\limits_{s=\pm}\frac{1}{\beta_s^3}\Bigg\{
\frac{8}{3}\beta_s-4\arctanh\beta_s
+\ln\bigg|\frac{4\mu^2-\omega^2(1+\beta_s)^2}{4\mu^2-\omega^2(1-\beta_s)^2}\bigg| \nonumber  \\
+\frac{\omega^2}{12\mu^2}&\sum\limits_{t=\pm
  1}\Bigg[t\Big(1+2t\beta_s\Big)\Big(1-t\beta_s\Big)^2\ln\bigg|\frac{4\Gamma^2(1-t\beta_s)^2}{4\mu^2-\omega^2(1-t\beta_s)^2}\bigg| \nonumber \\
&-\frac{2\mu}{\omega}
\Big(\frac{4\mu^2}{\omega^2}+ 3-3\beta_s^2)
\ln\bigg| \frac{2\mu-t\omega(1+t\beta_s)}{2\mu+t\omega(1+t\beta_s)}\bigg|\Bigg]\Bigg\},
\end{align}
\noindent which is valid 
for $|\beta_{\pm}| < 1$. Below, we show that
over a range of frequencies,
 $\epsilon_{zz}$ and $\epsilon_{\parallel}$
 can be tuned to achieve an ENZ response at
 different $\omega$.
 We also show that 
 there are a range of frequencies where
 $\epsilon_{\parallel}$ and $\epsilon_{zz}$
  can have opposite signs, which 
  strongly affects
 the types of modes that will propagate
 along the waveguide interface.
 Here we have neglected the effects of electron scattering on the absorption. 
 Increasing the dissipation in the diagonal components of the permittivity tensor can result in the generation of 
 other types of waveguide modes, including leaky waves \cite{leaky}, which fall outside the scope of this paper.
 
The remaining off-diagonal term $\gamma$
can play an important role in changing the polarization state of
 EM waves interacting
 with the WS via Faraday and Kerr rotations \cite{karg,Kotov1}.
Variations in the gyrotropic term can also cause shifts in
the surface plasmon frequency \cite{berg}.
The gyrotropic term
can be written analytically 
 in the limit $\omega\rightarrow 0$:
\begin{subequations}
\label{gam_approx}
\begin{align}
&\gamma = \frac{\alpha}{\pi \omega}\bigg[2v_FQ -\sum_{s=\pm}\frac{s\mu}{2\beta_s} \bigg(\frac{1}{\beta_s} \ln\bigg|\frac{1+\beta_s}{1-\beta_s}\bigg|-2\bigg)\bigg], |\beta_s|\ll1.
\end{align}
\end{subequations}
Thus, for fixed $\beta$, and
  small  tilting,
 $\gamma$ is a linear function of $\mu$,
 declining as the chemical potential increases.
 For arbitrary  $\beta$ and $\omega$, 
 it is necessary to resort to numerics to determine
 $\gamma$  (see Appendix).

We now make 
use of the
expressions above governing
the relevant  components of
${\rttensor{\epsilon}}$
to 
demonstrate  that by properly tailoring the
EM response,
the waveguide  structure 
can effectively transport localized EM energy 
over a broad range of frequencies and tilting of the Weyl cones 
(  $0<\beta<1$), where we recall 
 $\beta_+=-\beta_-=\beta$.
When characterizing the nontrivial behavior  of $\rttensor{\epsilon}$ 
in the WS, there are several relevant 
parameters to consider, 
including the chemical potential, frequency of the EM wave, 
tilt of the Weyl cones,
and the node separation parameter $Q$ (which is taken to be positive).

\subsection{Guided wave modes and energy confinement}\label{subsec2}
\begin{figure}[ht] 
\centering
\includegraphics[width=0.8\textwidth]{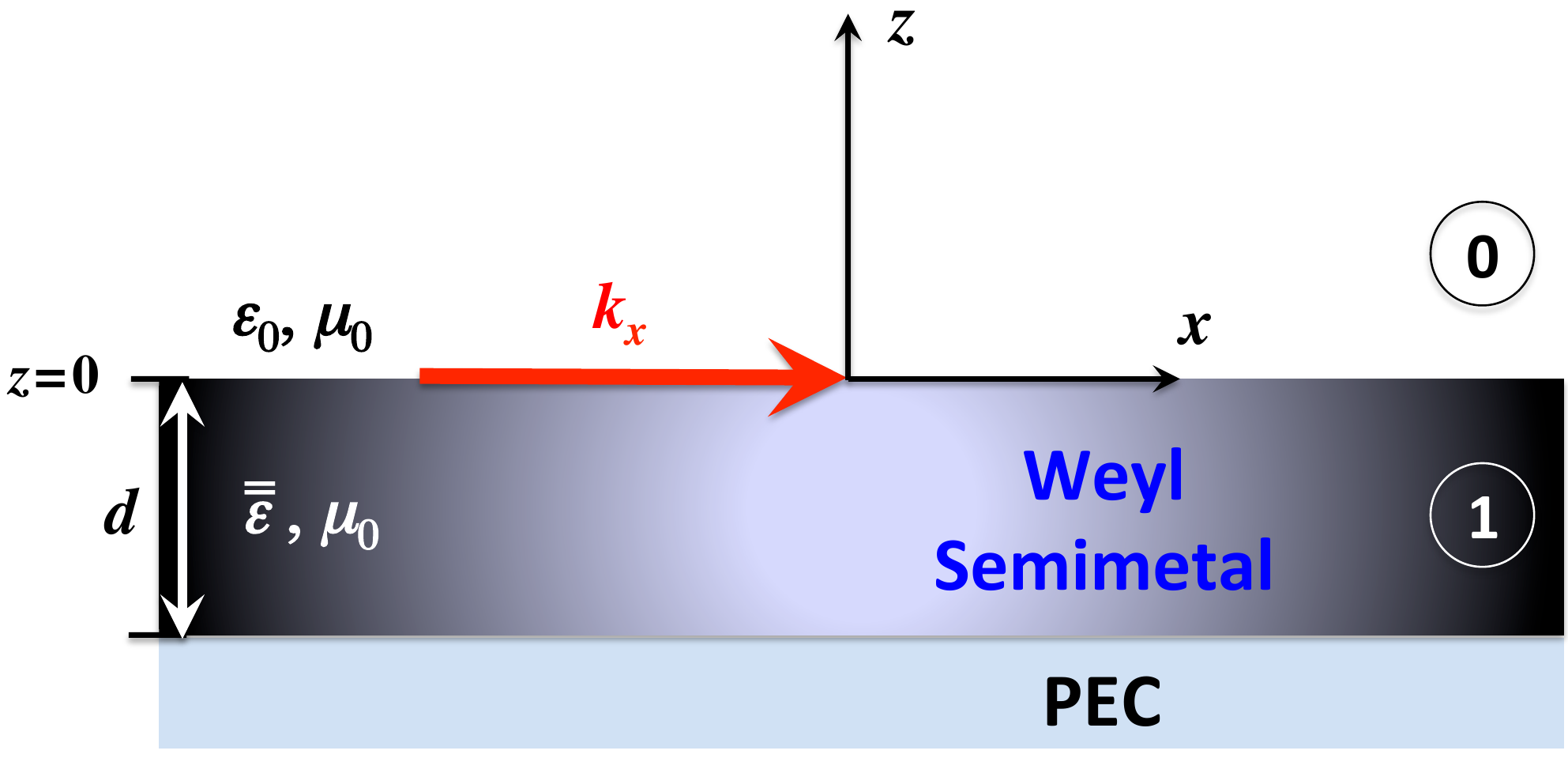}
\caption{Schematic of the
waveguide
 configuration
  involving a Weyl semimetal (in region $1$) with width $d$
  on top of 
a perfectly conducting substrate. In this configuration, the 
wavevector ${\bm k}$ resides in the plane of the WS (along $x$), while
the separation of Weyl nodes is perpendicular to the layers (along $z$).
The surrounding medium (region $0$) is taken to be vacuum.
}
\label{diagramxz}
\end{figure} 
We now 
investigate the 
EM modes 
for 
the  configuration shown in 
 Fig.~\ref{diagramxz}, which 
 consists of a planar
 Weyl semimetal (region $1$) adjacent to 
  a metallic substrate with perfect conductivity (PEC).
The propagation constant $k_x$ is invariant across each layer.
For both regions $0$ and $1$, we  implement   
Maxwell's equations for time harmonic fields,
\begin{subequations}
\begin{align} {\bm \nabla} \times {\bm E}_i &= i \omega \mu_0 {\bm H}_i, \label{h1} \\
{\bm  \nabla} \times {\bm H}_i &= -i\omega {\bm D}_i,
  \label{d1}
\end{align}
\end{subequations}
where $i=0$ or $1$.
Within the WS, the propagation vector ${\bm k}_1$ replaces the spatial derivatives,
transforming  Maxwell's equations into
 the forms, ${\bm k}_1 \times {\bm E}_1= \omega \mu_0 {\bm H}_1$ and
${\bm k}_1 \times {\bm H}_1 = -\omega {\rttensor{\epsilon}} \epsilon_0  {\bm E}_1$. 
These two equations together result in the following  expression for the ${\bm E}_1$ 
field in $\bm k$-space:
\begin{align}\label{wave}
{\bm k}_1 \times ({\bm k}_1 \times {\bm E}_1) = -k_0^2  {\rttensor{\epsilon}}  {\bm E}_1. 
\end{align}
 Using ${\bm k}_1 = k_{x} \hat{\bm x}+k_{1z} \hat{\bm z}$, and the identity  
${\bm k}_1 \times ({\bm k}_1 \times {\bm E}_1) = {\bm k}_1({\bm k}_1 {\bm E}_1) - k_1^2 {\bm E}_1 $,
permits expansion of  Eq.~(\ref{wave}),
\begin{align}\label{dmatrix}
\begin{pmatrix} 
k_0^2\epsilon_{xx}-k_{1z}^2  &i k_0^2\gamma &k_{1z} k_{x} \\
i k_0^2 \gamma&k_{\perp}^2- k^2_{0}\epsilon_{yy}& 0 \\
k_{x} k_{1z}  & 0  &k_0^2\epsilon_{zz}-k_{x}^2
\end{pmatrix} 
\begin{pmatrix}
E_{x1} \\
E_{y1} \\
E_{z1} 
\end{pmatrix}= 0,
\end{align}
where $k_0=\omega/c$, and 
$k_{\perp}=\sqrt{k_{1z}^2+k_{x}^2}$.
 Due to translational invariance, we also have $k_{1x}=k_{x}$.
The presence of all three $\bm E$ field components in Eq.~(\ref{dmatrix})  is
a consequence of the centrosymmetrc nature of the WS that can couple different 
polarization states, 
depending on
the WS material parameters and layer width.

Taking the determinant of the matrix in
Eq.~(\ref{dmatrix}) 
and setting it equal to zero,
gives the dispersion equation  for the wavevector  $k_{1z}$ in the semimetal:
\begin{align} \label{disp2p}
(\epsilon_{xx} k_0^2-k_{\perp}^2)(\epsilon_{xx}\epsilon_{zz} k_0^2-
\epsilon_{xx} k_{x}^2 -\epsilon_{zz} k_{1z}^2)
+k_0^2(k_{x}^2-\epsilon_{zz} k_0^2)\gamma^2=0.
\end{align}
Solving  Eq.~(\ref{disp2p})  for $k_{1z}$  results in two types of solutions
 denoted by $ k_+$ and $k_-$:
\begin{align} \label{kpm}
 k_{\pm} = \pm\sqrt{\frac{k_0^2}{2\epsilon_{zz}}\Biggl(
2  \epsilon_{zz}\epsilon_\parallel-(\epsilon_\parallel+\epsilon_{zz})
\kappa_x^2 
\pm \sqrt{(\epsilon_\parallel-\epsilon_{zz})^2 \kappa_x^4+
4\epsilon_{zz}\gamma^2 (\epsilon_{zz}-\kappa_x^2)}\Biggr)},
\end{align}
where $\kappa_x=k_x/k_0$.
The dispersion equation (\ref{disp2p}) 
can now be compactly written in terms of the two types of waves:
\begin{equation}
 \epsilon_{zz}(\kappa_{1z}^2-\kappa_+^2)(\kappa_{1z}^2-\kappa_-^2)=0,
\end{equation}
where $\kappa_{\pm} = k_{\pm}/k_0$, and  $\kappa_{1z} = k_{1z}/k_0$.

 When considering the transmission of EM surface waves, 
 the traveling
wave is confined to be within, or adjacent to, the waveguide walls.
For the configuration shown in Fig.~\ref{diagramxz}, where the $\var{x-y}$ plane is
translationally invariant,  we consequently  search for modes
that are localized around the interface separating  the WS and vacuum.
As with all open waveguide structures, the WS waveguide spectrum consists of a finite discrete set 
of guided modes with purely real longitudinal propagation 
constants and an infinite continuum of radiation modes. 
Although outside the scope of this work, 
leaky waves\cite{leaky} can also arise, which have  discrete complex 
longitudinal propagation constant solutions to the dispersion equation, 
and that  
decay longitudinally but do not obey the transverse radiation condition.

The magnetic field components 
in the vacuum region, ${\bm H}_0$, are thus written accordingly:
\begin{subequations}
\begin{align} 
H_{x0} &= r_3 e^{-k_{0z} z} e^{i k_{x} x}, \label{hobag1} \\
H_{y0} &= r_1 e^{-k_{0z} z}e^{i k_{x} x},\label{hobag2}\\
H_{z0} &= r_2 e^{-k_{0z} z} e^{i k_{x} x},\label{hobag3}
\end{align}
\end{subequations}
where $k_x$ is the propagation constant, $k_{0z}=\pm\sqrt{k_x^2-k_0^2}$,
and the coefficients $r_{1,2,3}$ are to be
determined upon application of the boundary conditions. 
Using the
Maxwell equation
${\bm\nabla\cdot}{\bm H}_0=0$,
a simple relation is found to exist
between the coefficients $r_2$ and $r_3$, namely:
$ r_3=-i r_2 {k_{0z}}/{k_{x}}$.
From  the magnetic field components above, we can use Eq.~(\ref{d1}) 
to easily deduce
the electric field components for the vacuum region:
\begin{subequations}
\begin{align} 
E_{x0} &= Z_0 r_1 i \frac{k_{0z}}{k_0}e^{-k_{0z} z} e^{i k_{x} x}, \label{eobag1} \\
E_{y0} &= Z_0 r_2 \frac{k_0}{k_x} e^{-k_{0z} z}e^{i k_{x} x},\label{eobag2}\\
E_{z0} &= -Z_0 r_1 \frac{k_x}{k_0} e^{-k_{0z} z} e^{i k_{x} x},\label{eobag3}
\end{align}
\end{subequations}
where $Z_0=\sqrt{\mu_0/\epsilon_0}$ is the impedance of free space.

For the WS region, when implementing  Maxwell's equations,
the anisotropic nature of the WS must be explicitly taken into account by involving Eq. (\ref{rttepsilon}) in the calculations.
Due to the $\pm$ signs in Eq.~(\ref{kpm}),
the general solution to the ${\bm E}$ field in the WS region is
a linear combination
of the  four branches of the wavevector  $k_{1z}=\{k_+,-k_+,k_-,-k_-\}$:
\begin{align} \label{ey1}
E_{y1}=(a_1 e^{i k_+ z}+a_2e^{-i k_+ z}+a_3 e^{i k_- z}+a_4 e^{-i k_- z})e^{i k_{x} x}.
\end{align}
To determine the set of coefficients $\{r_1,r_2,r_3\}$, and $\{a_1,a_2,a_3,a_4\}$, it is necessary to invoke 
matching interface conditions and  
boundary conditions.
But first,  the remaining $\bm E$ and $\bm H$ fields must be constructed.
This can be accomplished by
first using Eq. (\ref{d1})  to  write  the following relations:
\begin{subequations}
\begin{align}
&\frac{\partial H_{y1}}{\partial z}=i\omega \epsilon_0(\epsilon_\parallel E_{x1}+i\gamma E_{y1}),\\
&\frac{\partial H_{x1}}{\partial z}-ik_{x} H_{z1}=i\omega \epsilon_0 (i\gamma E_{x1}-\epsilon_\parallel E_{y1}), \label{emid} \\
&k_{x} H_{y1}=-\omega \epsilon_0 \epsilon_{zz} E_{z1}.
\end{align}
\end{subequations}
While the other Maxwell equation,  Eq.~(\ref{h1}), gives,
\begin{subequations}
\begin{align}
&\frac{\partial E_{y1}}{\partial z}=-i\omega\mu_0 H_{x1}, \label{hx1} \\
&\frac{\partial E_{x1}}{\partial z}-ik_{x} E_{z1}=i\omega\mu_0 H_{y1}, \label{hmid}\\
&k_{x} E_{y1}=\omega \mu_0 H_{z1}. \label {hz1}
\end{align}
\end{subequations}
Here the translational invariance in
the $x$-direction has been used for the derivatives: $\partial_x \rightarrow i k_{x}$.
Inserting Eq.~(\ref{ey1}), into the equations above,
it is now possible to write each component of the EM field in terms of the coefficients $\{a_1,a_2,a_3,a_4\}$.
For example, $H_{x1}$ and $H_{z1}$ are easily found from Eqs.~(\ref{hx1}) and (\ref{hz1}) respectively. From that,
one can solve Eq.~(\ref{emid}) for
$E_{x1}$, and so on. Explicit details 
regarding the coefficients can 
be found in Appendix \ref{appdxB}.

Upon matching the tangential electric and magnetic fields at the vacuum/WS interface, and 
using the boundary conditions of vanishing tangential electric fields at the ground plane,
it is straightforward to determine the characteristic equation 
that governs the EM modes of our structure.
The resultant transcendental equation that must be satisfied 
is written in the following form:
\begin{align}
\label{bigD}
 &-\kappa_+ \cos q_+ \sin q_- \Big\{ \kappa_x^2 ( \epsilon_{zz}-1)  ( \kappa_x^2-\epsilon_\parallel ) + \epsilon_{zz}  (\kappa_-^2 -\kappa_+^2 ) +  \kappa_x^2 (  \epsilon_{zz} \kappa_+^2-\kappa_-^2) \Big\} \nonumber \\
 &+\kappa_- \cos q_- \sin q_+ \Big\{ \kappa_x^2( \epsilon_{zz}-1)  ( \kappa_x^2-\epsilon_\parallel )+ \epsilon_{zz}  ( \kappa_+^2 -\kappa_-^2 ) +  \kappa_x^2 (  \epsilon_{zz} \kappa_-^2-\kappa_+^2)\Big\} \\ \nonumber 
 & +
 \epsilon_{zz}   \kappa_+ \kappa_-  \kappa_{0z} \cos q_- \cos q_+(\kappa^2_- - \kappa^2_+)+
\kappa_{0z}  \sin q_+ \sin q_- ( \kappa_x^2-\epsilon_{zz} )  (\kappa^2_- - \kappa^2_+)=0.
\end{align}
Here we have introduced the dimensionless 
quantities: $q_{\pm}~=~k_{\pm} d$, and
$\kappa_{0z}=k_{0z}/k_0$. 
For the lossless guided wave modes studied in this paper,
 the dispersion equation (Eq.~(\ref{bigD})) is either purely imaginary or real.
In the absence of gyrotropy ($\gamma=0$), Eq.~(\ref{bigD}) becomes decoupled, and
the corresponding dispersion equation for purely diagonally
anisotropic media reduces to \cite{hmm},
\begin{align}
   \tan \Big(k_0 d \sqrt{\epsilon_\parallel -\kappa_x^2}\Big)&=-\frac{\sqrt{\epsilon_\parallel -\kappa_x^2}}{\sqrt{\kappa_x^2-1}},  \label{spol} 
   \end{align}
   and,
   \begin{align}
 \tan \Big(k_0 d \sqrt{\epsilon_\parallel (1-\kappa_x^2/\epsilon_{zz})}\Big)&=\frac{ \sqrt{\epsilon_\parallel(\kappa_x^2-1)}}
 {\sqrt{1-\kappa_x^2/\epsilon_{zz}}}. \label{ppol}.
\end{align}
The allowed modes in Eq.~(\ref{spol}) depend only on the in-plane component of the permittivity $\epsilon_\parallel$,
and correspond to a TE  polarized
state where the EM wave only has components ($E_y,H_x,H_z$).
The dispersion in Eq.~(\ref{ppol}) depends on both $\epsilon_\parallel$ and $\epsilon_{zz}$, and
corresponds to a TM 
polarized state with components ($E_x,E_z,H_y$).
Due to the dependence on both
diagonal permittivity components, 
and the potential for highly  localized guided waves, 
the TM 
state is naturally of greater interest.
When the off-diagonal elements in $\rttensor{\epsilon}$ are negligible, 
as can occur e.g., when  $d/\lambda \ll 1$,  Eqs.~(\ref{spol})-(\ref{ppol})
accurately account for the guided mode solutions. 
When the 
gyrotropic parameter is negligible, 
an incident EM wave 
interacting with the WS retains its initial 
polarization state, thus limiting complications 
from Faraday and Kerr effects.
Of course, as discussed above,  
when  $\gamma$ is not negligible, 
all three polarization directions must be taken into account,
and nontrivial coupling can arise between the components of 
the ${\bm E}$
and ${\bm H}$ fields.

An insightful  quantity that
yields the fraction of EM energy that is confined
 to the WS region is
the confinement factor $\eta$, 
defined as,
\begin{align}
\eta=\dfrac{1}{1+{\cal U}_0/{\cal U}_{1}},
\end{align}
where 
 ${\cal U}_i\equiv \int_{\Omega_i}  dz\, {U}_i(z)$ is 
the energy density ${U}_i$ integrated over  region $\Omega_i$ ($i=0,1$).
The energy density,
which accounts for any possible anisotropy and
dissipation present,  is  written,
\begin{align}
{U}_i(z)=\frac{1}{4}\Re{\left[ {\bm E}_i^\dagger \cdot \frac{\partial(\omega  \rttensor{\epsilon}_i)}{\partial \omega} {\bm E}_i+
{\bm H}_i^\dagger \cdot \frac{\partial (\omega \mu_i)}{\partial \omega} {\bm H}_i\right]},
\end{align}
which ensures that the energy density is positive, 
as required by causality.

Related to the energy density, is the 
time-averaged energy flow, given by the real part of the
Poynting vector $\bm S$, 
which for energy  flow along the direction of the interface is
expressed as,
\begin{equation}
S_{xi}(z) = \frac{1}{2} \left( E_{yi} H^*_{zi}-E_{zi} H^*_{yi} \right),\;\;\;  \text{for } i=0,1.
\end{equation}
The corresponding power in each region $P_{xi}$ is
then 
defined in the usual way as an integral
of the Poynting vector over
region $i$: $P_{xi}=\int_{\Omega_i} dz S_{xi}$.
From the Poynting vector and energy density defined above,
it is possible to construct another 
physically relevant quantity, namely  the energy transport velocity 
${\bm v}_T$ \cite{vt1,vt2},
which is the velocity at which EM energy is transported
through a given region of the waveguide structure, and
defined as:
\begin{align}
{\bm v}_T\equiv \frac{\langle{\bm S}\rangle}{\langle U\rangle},
\label{vtavg}
\end{align}
where the 
brackets indicate spatial averaging 
over all space, and 
only time-averaged quantities are considered here.

\subsection{Discussion and results} \label{subsec3}
Having established the methods for determining the waveguide modes for
the WS structure, 
we now consider a range of material and geometrical parameters
that leads to the propagation of localized EM waves along the WS surface.
We consider subwavelength widths $d/\lambda \ll 1$, leading to a decoupling of the
EM field components  and excitation of TM modes where the components $E_x,E_z$, and $H_y$
dominate.
In cases where light localization and concentration is extremely high, 
the Fermi arc surface states can play an important role in 
the dispersion of surface plasmon-polaritons \cite{qchen}.
We search for modes that are in the parameter regime where 
the dissipative components of $\epsilon_{zz}$ and $\epsilon_{\parallel}$ vanish.
From the previous discussion in Sec.~\ref{subsec1}, this implies 
that we focus on frequency bands that satisfy 
$\omega< 2\mu/(1+\beta)$. 
The process for calculating the modes is as follows:
For a given frequency  $\omega$,
and relevant WS  parameters,
each component of the tensor $\rttensor{\epsilon}$
is calculated. These components are inserted
into the transcendental equation (\ref{bigD}) to
determine the permitted propagation constants $\kappa_x$
that are associated with the chosen $\omega$.
This creates a map of a finite number of dispersion bands
for a given waveguide width $d$ that
ensure the boundary conditions are fully satisfied.
Finally, from the dispersion diagram, 
the EM fields, and energy characteristics can
be straightforwardly extracted.
In presenting the results, we generally scale  $\omega$ by the energy
 unit $v_F |Q|$, and define $\widetilde{\omega}\equiv \omega/(v_F |Q|)$.
 Similarly, when considering the chemical potential in the WS, 
we take the representative 
 dimensionless value corresponding to  $\widetilde{\mu}\equiv\mu/(v_F |Q|)=0.2$.
 We also
 set the energy cutoff to $\widetilde{\Gamma}\equiv \Gamma/(v_F |Q|) \sim 8$,
so that the linearized model remains applicable.

\begin{figure}[ht] 
\centering
\includegraphics[width=\textwidth]{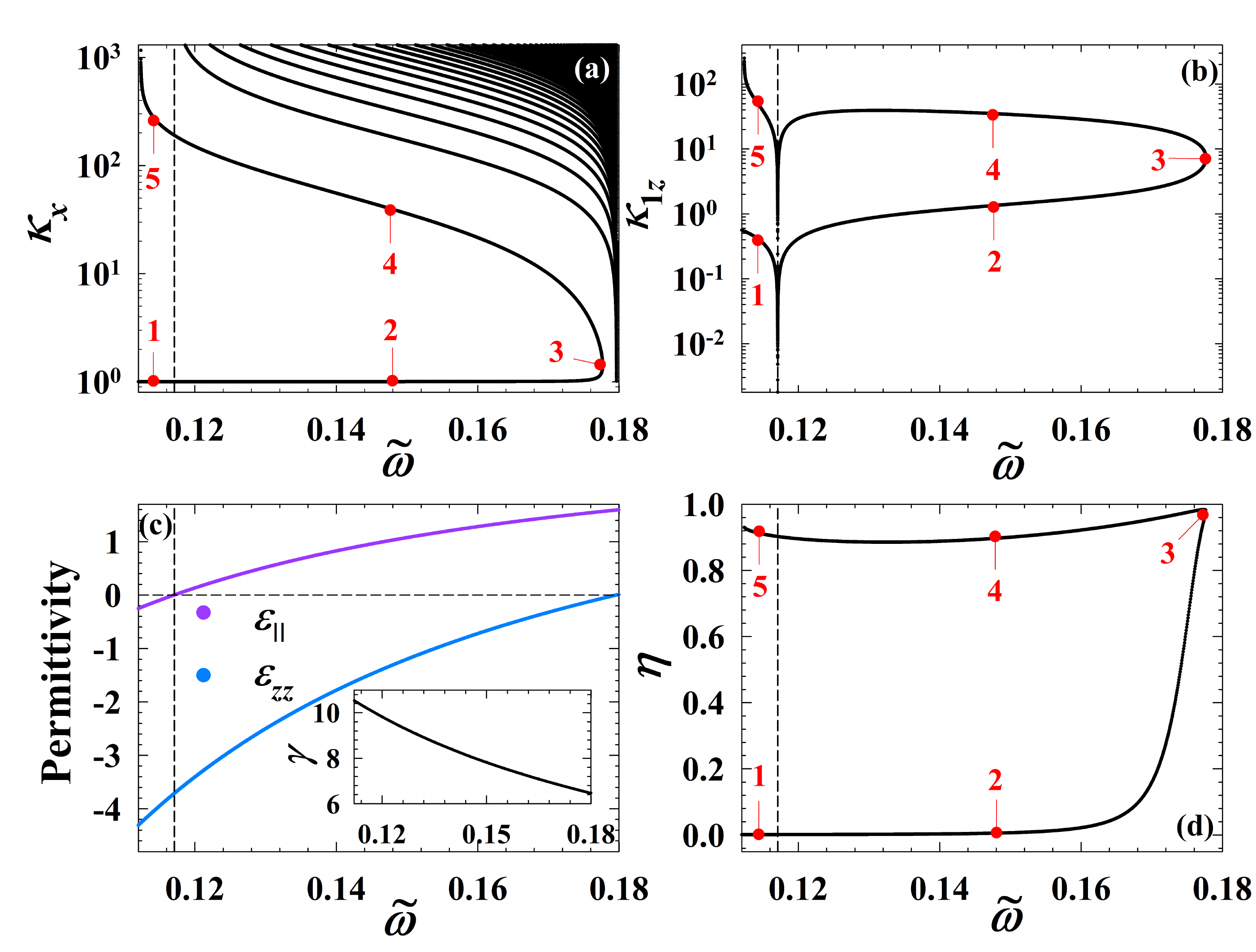}
\caption{
Diagrams depicting 
the scaled frequency dependence to the
relevant physical quantities for the Weyl semimetal 
waveguide shown in Fig.~\ref{diagramxz}. 
(a) The mode diagram showing the dimensionless propagation  
constant $\kappa_x$. 
The numbers $1\rightarrow 5$ identify regions of the diagram discussed below.
(b) The normalized wavevector in the WS region $\kappa_{1z}$
for the labeled band in (a).
(c) The frequency dependence of the
diagonal permittivity components $\epsilon_{\parallel}$ and $\epsilon_{zz}$ are shown in the main plot
while the off-diagonal component $\gamma$ is shown in the inset.
(d) The energy confinement factor $\eta$ for the modes calculated in (a).
Note that $\kappa_x$ and $\kappa_{1z}$ are shown on a logarithmic scale.
The modes to the left of the vertical line  in panel (a) correspond to 
the frequency range in which $\epsilon_{\parallel}$ and $\epsilon_{zz}$
are both negative [see panel (c)]. 
The dimensionless WS material parameters are set to $\beta=0.9$ and 
$\widetilde{\mu}=0.2$. 
}
\label{dispvsomega}
\end{figure} 
%
%
In Fig.~\ref{dispvsomega}(a),
we present the dispersion diagram 
in terms of the dimensionless propagation constant $\kappa_x$ and frequency $\widetilde{\omega}$.
Various points are numerically labeled on the diagram for discussion below.
We consider subwavelength widths corresponding to $k_c d = 0.03$,
where $k_c$ is a characteristic wavenumber corresponding to $k_c=\omega_c/c$,
at the  normalized  design frequency $\widetilde{\omega}_c=0.12$.
The conical tilt is fixed at $\beta=0.9$.
As seen, there are a discrete number of bands,
which arise from the geometrical constraint placed upon the fields from the bounding
surfaces. 
Also, since the fields in the vacuum decay over the length scale $k^{-1}_0$,
regions of the dispersion curves near the light line ($\kappa_x \sim 1$) 
correspond to a larger decay length,
and thus the fields reside mainly outside the guiding layer.
The curves that lie to the left of the vertical dashed line
identify 
modal solutions 
that are predominately evanescent,
with the EM field decaying on both sides of the WS/vacuum interface. 
This arises when
both $\epsilon_{\parallel}$ and $\epsilon_{zz}$ are negative.
For frequencies  to the right of the vertical line,
the modes
 transition from predominately evanescent in the WS,
to sinusoidal,
whereby
 $\epsilon_{\parallel}>0$ and $\epsilon_{zz}<0$.
The modes at these frequencies of course 
still decay in the vacuum region.
In Fig.~\ref{dispvsomega}(b),
the normalized wavevector 
in the WS,  $\kappa_{1z}$, is shown.
Due to the decoupling of the TE and TM modes,
we have   
$\kappa_\pm \rightarrow \kappa_{1z}$.
At the transition point between 
oscillatory and evanescent 
modes (vertical line at $\widetilde{\omega} \approx 0.117$),
$\epsilon_{\parallel}\approx 0$, and consequently the 
wavevector drastically diminishes, corresponding to 
a very long wavelength in the WS.
On the other hand, when $\epsilon_{zz}\approx0$
(point 3), the wavevector is noticeably larger.
The correlations between the 
mode diagrams  and the EM response
is shown in Fig.~\ref{dispvsomega}(c).
Notably, 
$\epsilon_{\parallel}$ and $\epsilon_{zz}$
become vanishingly small at different frequencies,
leading to an interesting anisotropic ENZ response.
 In particular, at the crossover 
  point $\widetilde{\omega} \approx 0.117$, 
   $\epsilon_{\parallel}\approx 0$, and $\epsilon_{zz}<0$.
   Indeed, when the energies associated with the chemical potential and frequency 
   are similar,
   $\epsilon_{\parallel} \approx 0$ when 
   $\omega\approx 2 \mu/\sqrt{3 \pi/\alpha + 
 \log(4 \Gamma^2/(3\mu^2))}$.
 The other ENZ scenario occurs at the higher 
  frequency $\widetilde{\omega} \approx 0.18$,
 where
 $\epsilon_{zz}\approx 0$, and $\epsilon_{\parallel}>0$.
This is also in the vicinity  where the slope of the
 dispersion curve in Fig.~\ref{dispvsomega}(a) changes  sign (labeled 3). 
This feature of the dispersion 
diagram corresponds to where 
 $\kappa_x$ close to the light line bends back
 and rapidly increases.
Similar  behavior is seen in general, when a waveguide
is in contact with another medium possessing 
permittivity components that are opposite in
sign.
The 
corresponding energy flow along  the interface
that separates the two media
undergoes an
abrupt reversal when going from vacuum to the WS or vice versa.
Additionally, extreme field enhancement can occur when either 
of the permittivity components is near the ENZ regime. 
For example, since
the boundary conditions dictate that the ratio
of the normal components of the electric fields 
at the semimetal interface are $E_{z1}/E_{z0}\sim1/\epsilon_{zz}$,
a substantial mismatch in field strengths can clearly occur when $\epsilon_{zz} \sim 0$.
Although its effects are weak for a subwavelength WS,
for completeness
the inset presents 
the frequency dependence to the off-diagonal component $\gamma$
calculated via the methods described in the Appendix.
Lastly, in Fig.~\ref{dispvsomega}(d),
the energy confinement factor $\eta$
is shown as a function of normalized frequency. 
The labeled numbers identify  the 
corresponding  ($\kappa_x$, $\widetilde{\omega}$) pairs in Fig.~\ref{dispvsomega}(a).
The fraction of time-averaged 
EM energy within the WS at frequency points 1 and 2
is shown to be extremely small 
due to the fact that
modes with $\kappa_x \sim 1$ 
correspond to the EM fields residing mainly in the vacuum region.
The upper branch at points $3 \rightarrow 5$ reveals that
a substantial percentage of the energy density is contained within 
the semimetal. We see that $\eta$ is maximal at position 3, 
where the slope of the dispersion diagram
changes sign [see Fig.~\ref{dispvsomega}(a)], and $\epsilon_{zz}\approx 0$, giving rise to a large
increase of the
electric field normal to the interface.

\begin{figure}[t] 
\centering
\includegraphics[width=\textwidth]{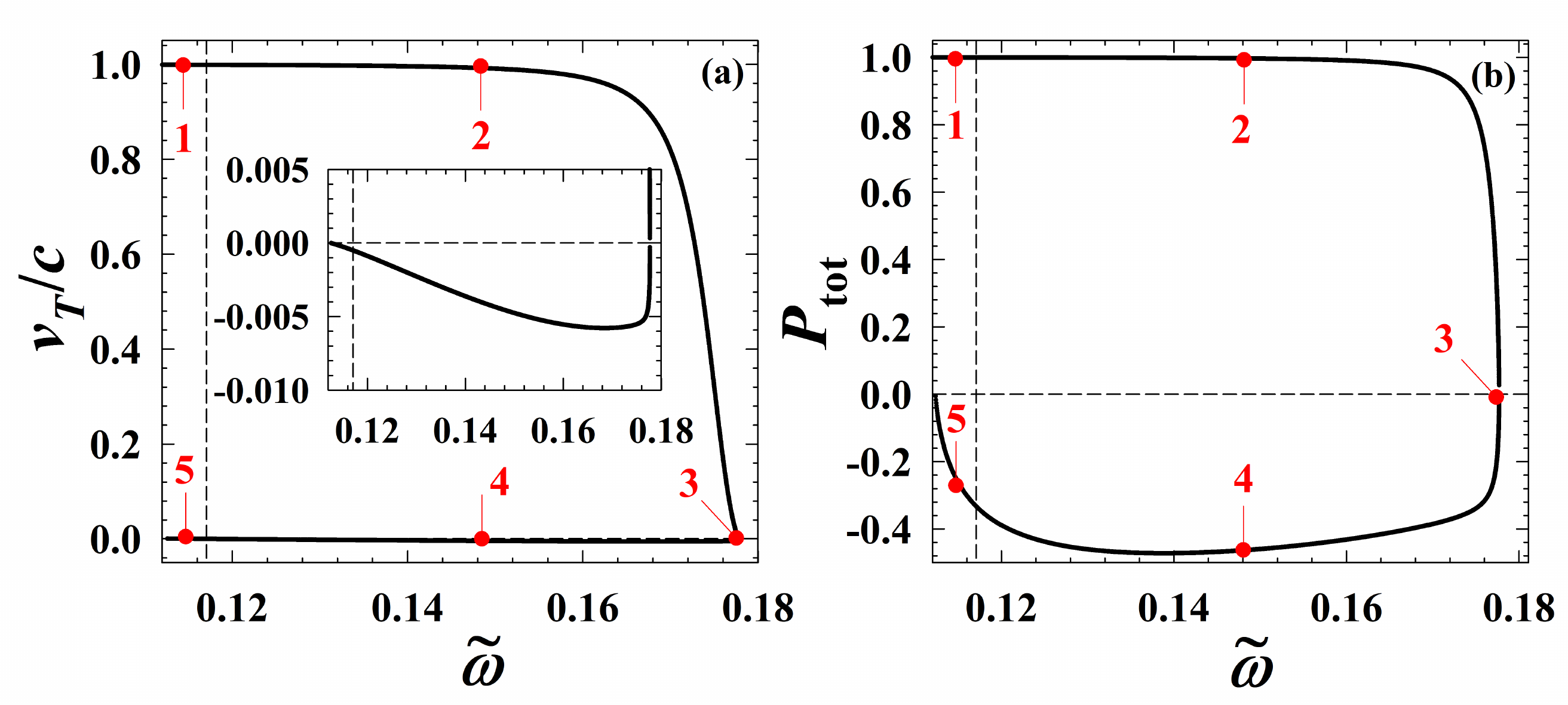}
\caption{
(a) The energy transport 
velocity (normalized by the speed
of light $c$)
as a function of dimensionless frequency.
The inset is a magnification 
of  $v_T$ for the lower curve along the points
$3\rightarrow5$.
(b) The total normalized
power of the system.
Both $v_T$ and $P_{\rm tot}$ are 
calculated for the modes
described by the first dispersion curve in Fig.~\ref{dispvsomega}(a),
and correlated with the number labels.
The WS has its conical
tilting parameter set to $\beta=0.9$.  
}
\label{power}
\end{figure} 
%
To gain additional perspectives on the energy flow
in the system,
we present
 in Fig.~\ref{power}(a),
the energy transport velocity  $v_T$
as a function of the normalized frequency. 
To calculate this quantity we take the ratio of the
Poynting vector to the energy density per Eq.~(\ref{vtavg}).
For the subwavelength structure under consideration, 
the time-averaged energy density can be  
calculated approximately for the WS region:
\begin{align}
U_1\approx \frac{\mu  (\kappa_x^2-1)  }{4 (\epsilon_{zz}-\kappa_x^2)^2 \sin^2 (k_+ d)}
\bigg\{(\epsilon_{zz}^2
&+\frac{\partial (\omega\epsilon_{zz})}{\partial \omega} \kappa_x^2) 
\kappa_+^2 \cos^2(k_+d [1+\frac{z}{d}])  \nonumber \\ 
&+\frac{\partial (\omega\epsilon_{\parallel})}{\partial \omega} 
(\epsilon_{zz}-\kappa_x^2)^2 \sin^2(k_+d[1+\frac{z}{d}])\bigg\},
\end{align}
\noindent 
while for the vacuum region, we have:
\begin{align}
 U_0 = \frac{\mu\kappa_x^2}{2} e^{-2 k_0 z \sqrt{\kappa_x^2-1} }.
\end{align}
\noindent Likewise, the time-averaged 
Poynting vector component $S_x$ (for $d/\lambda \ll 1$) can be expressed as,
\begin{align}
S_{x1}
\approx 
\frac{c \mu \epsilon_{zz}  \kappa_x (\kappa_x^2-1) \kappa_+^2 \cos^2(k_+d[1+ z/d])}
{2 (\epsilon_{zz}-\kappa_x^2)^2 \sin^2(k_+ d)},
\end{align}
and for the vacuum region:
\begin{align}
S_{x0}= \frac{c\mu \kappa_x}{2} e^{-2 k_0 z \sqrt{\kappa_x^2-1} }.
\end{align}
In the absence of loss, the energy transport direction lies
solely along the interface ($x$-direction), 
so that ${\bm v}_T = v_T \hat{\bm x}$.
Note that when $\kappa_x \rightarrow 1$,
$v_T$ corresponds to the expected
phase velocity, or velocity at which plane wavefronts travel
along the $x$-direction: 
$v_T={c}/{\kappa_x}$.
It is evident from Fig.~\ref{power}(a),  
that  $v_T$
obeys casualty, with $v_T\leq c$.
Due in part to the competing effects that can arise in
structures with counter-propagating energy 
flows on 
opposite sides of the waveguide interface,
the  transport of energy
can occur at effective speeds smaller than $c$.
In particular, it is evident that $v_T$ can become very small
for the modes labeled $3\rightarrow 5$.
Note that from the inset, 
we see that over the entire 
frequency range,
$v_T$ is slightly negative, but never zero until
at the crossover point (3), where the dispersion curve in Fig.~\ref{dispvsomega}(a)
changes direction.
For modes 1 and 2,  where $\kappa_x \approx 1$,
a large portion of the EM field resides in the vacuum region, 
and thus the net transport of energy has $v_T/c \approx 1$.
Slow-light phenomena have been explored 
previously \cite{metaslow,board} in ENZ and metamaterial systems.
The 
flow of energy involves both
the interplay between the propagating and stored  forms of energy.
The proposed  WS waveguide thus serves as an effective platform 
for studying slow-energy effects
in dissipationless systems, where the
energy propagation  in adjacent regions  
are oppositely directed due to the sign change
of the permittivity across the interface.
Having  tunable control over the speed 
of energy transport
can also have practical uses in optical 
memory devices, 
biosensing \cite{toca}, and chemical sensors \cite{chemy}.

As a complimentary study, 
we show in Fig.~\ref{power}(b),  the total power $P_{\rm tot}$ of the system, 
where  we define,
\begin{equation}
P_{\rm tot} = \frac{P_{x0}+P_{x1}}{|P_{x0}|+|P_{x1}|}.
\end{equation}
The upper branch of the power curve at unity (modes $1 \rightarrow 2$)
corresponds to the dispersion curve where $\kappa_x\sim 1$
in Fig.~\ref{dispvsomega}(a), where
 the energy flows predominately outside the semimetal.
 The net power
 drops drastically to zero at  the  transition point (mode 3),
 where the slope of the dispersion curve changes sign.
 This follows from the superposition of two counter-propagating 
 energy flows, and 
is consistent with the behavior of the 
group velocity $v_g\equiv \partial \omega/\partial k_x$, 
which becomes
vanishingly small at point 3.
Reducing the frequency, and going from points 
$3\rightarrow 5$,
it is apparent that
the power reverses
 and now has a net flow in the opposite (negative) direction,
corresponding to the upper branch of the dispersion curve in Fig.~\ref{dispvsomega}(a), 
until
the net power again vanishes at $\widetilde{\omega} \approx 0.112$. 

\begin{figure}[ht] 
\centering
\includegraphics[width=0.9\textwidth]{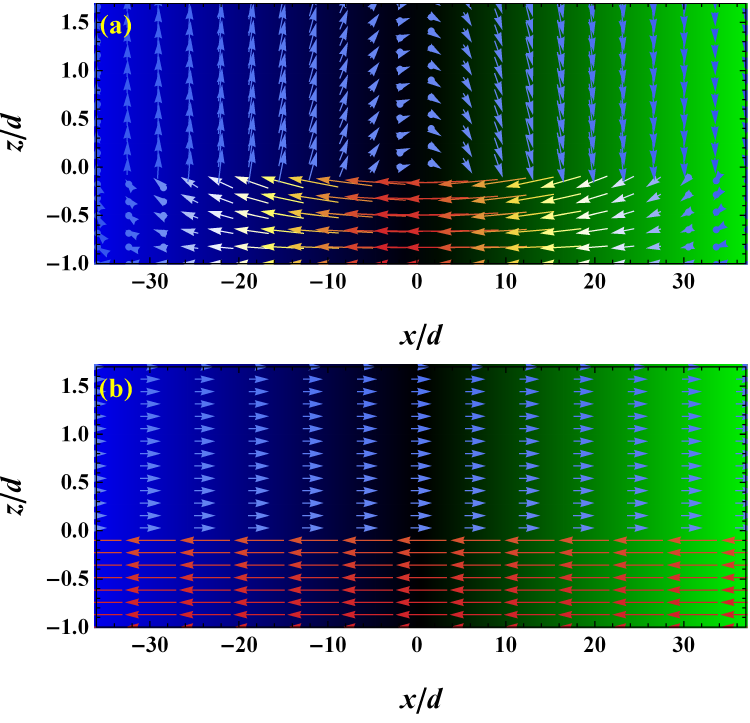}
\caption{
(a) Representative snapshot of the instantaneous Poynting vector ${\bm E} \times {\bm H}$.
(b) The time-averaged Poynting vector, averaged over one cycle.
The frequency in both cases
corresponds to the point on the dispersion diagram in Fig.~\ref{power}(a)
where the energy velocity vanishes (labeled 3). 
The interface between the vacuum region and the WS is located at $z=0$.
}
\label{poynt}
\end{figure} 
As alluded to earlier, for frequencies in which the diagonal permittivity  components  
change sign when crossing into
another medium,
the resultant 
flow of energy in each region can often times be oppositely directed.
To visualize the directional dependence to the energy flow, we present in Fig.~\ref{poynt}, 
the spatial behavior of the Poynting vector in the surrounding vacuum region ($z/d>0$), and within the WS
($0\leq z/d \leq -1$).
We consider the frequency and propagation constant corresponding to 
point 3 on the dispersion curve in Fig.~\ref{dispvsomega}(a). 
At this point $\epsilon_{zz}$ is slightly negative,
with $\epsilon_{zz} \approx 0$, and $\epsilon_{\parallel}\approx 1$.
First, in Fig.~\ref{poynt}(a), 
the instantaneous Poynting vector is shown.
We see that the energy flow tends to 
form a vortex-like pattern in which the flow of energy upward is
counteracted by the flow downward. Since the normal component of the ${\bm E}$ field 
is discontinuous at the interface, the energy flow
undergoes  an abrupt change at $z=0$.
Since there is no dissipation however, when averaging over a complete period,
there is no net power flow in the $z$ direction.
This is seen in Fig.~\ref{dispvsomega}(b), where the time-averaged Poynting vector is presented.
The cancellation of the energy flow normal to the interface leads to
 a net propagation of energy only along the interface ($x$ direction).
 We saw earlier in Fig.~\ref{power}(b) that this mode corresponds to a vanishing net power flow. 
 Thus, the   spatially integrated Poynting vector component  along the interface in each region
 completely cancel one another, leading to a net effective energy velocity of zero [see Fig.~\ref{power}(a)].

\begin{figure}[ht]
\centering
\includegraphics[width=\textwidth]{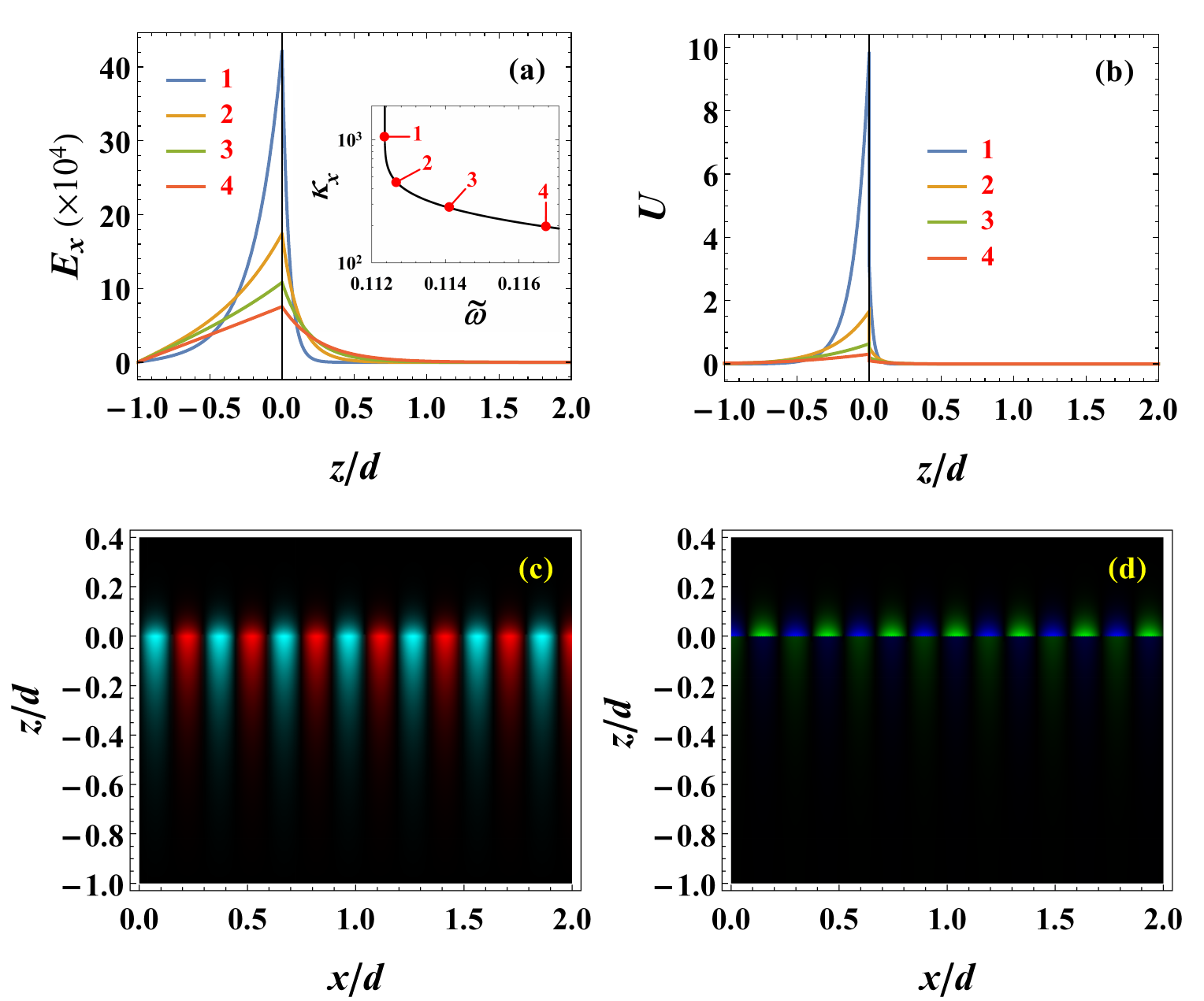}
\caption{
Spatial behavior of the electromagnetic field profiles.
(a)  The $x$-component of the electric field $E_x$
is plotted as a function of the normalized coordinate $z/d$. 
In (b) the 
electromagnetic 
energy density 
$U$ is shown. 
The legends in (a) and (b) identify the four frequencies considered along
the dispersion curve shown
in the inset.
As the frequency increases, the overall  field
amplitudes are shown to decline.
The vertical line at  $z=0$ locates the vacuum/WS interface.
Panels (c) and (d) are color maps depicting  
$E_{x}$ and $E_{z}$, respectively. Bright regions
correspond to high field intensities.
The frequency corresponds to the mode labeled 1 in panel (a),
and the tilt parameter is set to $\beta=0.9$.
}
\label{fields2}
\end{figure} 
%
%
We now turn our attention to the explicit behavior of the fields themselves.
In Fig.~\ref{fields2}(a) we show the spatial behavior of the  electric field 
parallel to the interface $E_x$. Four representative frequencies are considered 
along the dispersion curve show in the inset. These modes correspond to the
regime that is left of the vertical dashed line in Fig.~\ref{dispvsomega}(a), and
thus these modes are strictly evanescent on both sides of the interface.
It is noted that as  the frequency 
approaches the lowest frequency case, $\widetilde{\omega}\approx 0.112$
(labeled 1),
the electric field becomes strongly enhanced.
As was observed in Fig.~\ref{power}, this correlates with
the frequency in which $v_T\approx 0$, and the slope of the dispersion curve vanishes.
Note that over the considered  frequency range,
both diagonal components 
of $\rttensor{\epsilon}$ are negative, and $\epsilon_{\parallel} \approx 0$.
Thus, these are purely evanescent modes with
field profiles that decay
on both sides of the vacuum/WS interface,
as is clearly seen in Fig.~\ref{dispvsomega}(a).
The energy density, which contains both the electric and magnetic fields
also exhibits strong field localization at the interface as seen Fig.~\ref{dispvsomega}(b). 
Next in Figs.~\ref{fields2}(c) and \ref{fields2}(d), color maps illustrate the full spatial profiles 
the tangential ($E_x$) and normal ($E_z$) electric field components, respectively. 
Here the propagating term $e^{i k_x x}$ is included for all field components.
The tilt parameter $\beta$ is set to $\beta=0.9$, while  $\widetilde{\omega}$
is fixed to the frequency labeled 1 in the mode diagram shown in the inset of Fig.~\ref{fields2}(a).
It is evident from  Figs.~\ref{fields2}(c) and \ref{fields2}(d) that
the propagating electric field has both  field polarizations 
strongly confined  to the interface. 
 As shown  in Figs.~\ref{fields2}(a) and \ref{fields2}(b), the  EM fields
 also become substantially enhanced at this frequency. 

\begin{figure}[ht]
\centering
\includegraphics[width=\textwidth]{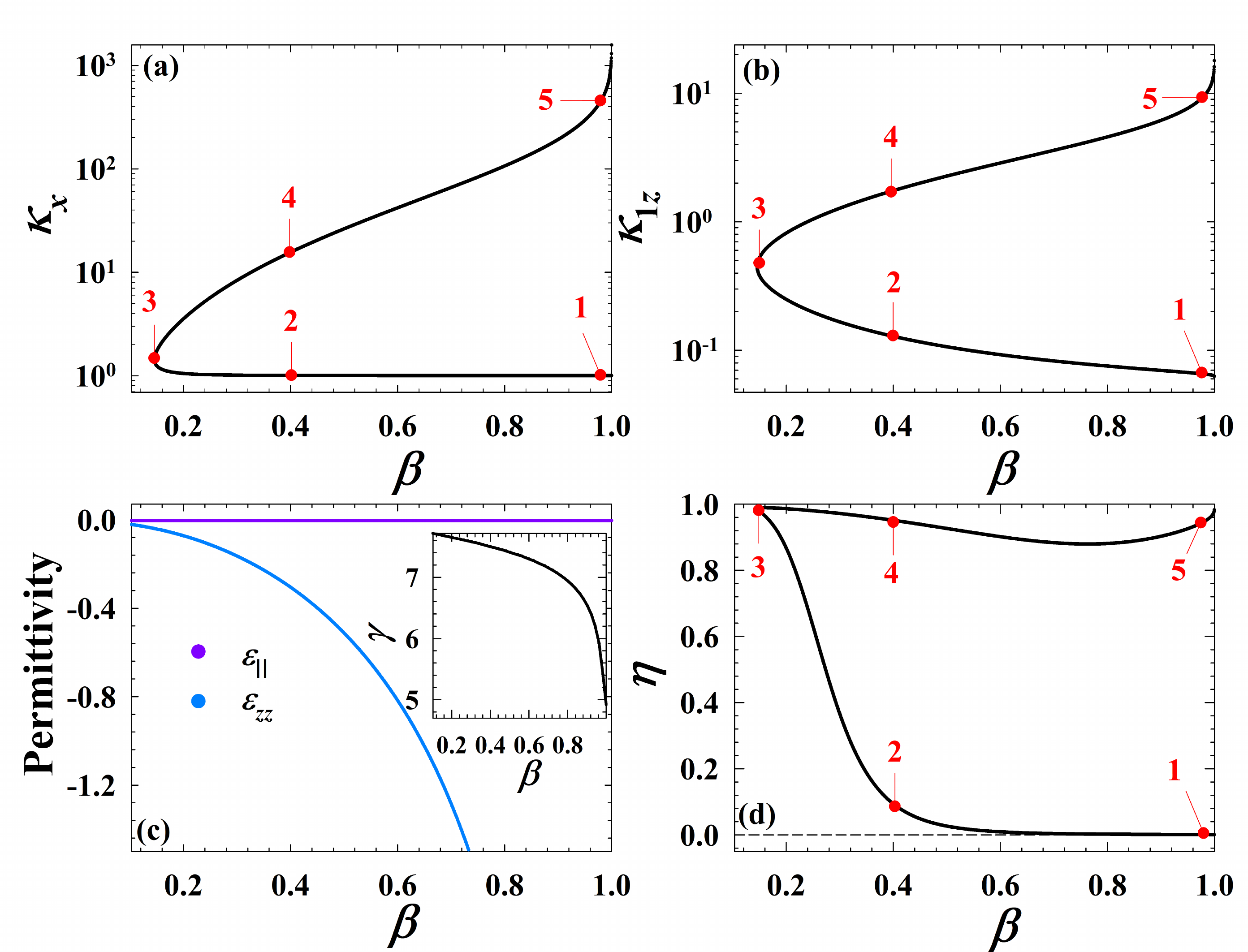}
\caption{
The  effects of the conical tilt $\beta$ on the waveguide characteristics.
In (a) the propagation constant $\kappa_x$ is shown vs $\beta$.
The wavevector in the WS, $\kappa_{1z}$, is shown in (b).
The electromagnetic response 
of the WS waveguide is shown in (c), where the diagonal components of the permittivity tensor
$\epsilon_{\parallel}$ and $\epsilon_{zz}$ are presented in the main plot, while the inset reveals
the off-diagonal component $\gamma$. 
The  fraction of EM energy $\eta$ that is confined to the 
semimetal for the modes calculated in (a) is shown in (d). 
The normalized frequency corresponds to
$\widetilde{\omega}=0.117$, which is the point labeled 5 of the dispersion curve
shown in Fig.~\ref{dispvsomega}(a). Note that $\kappa_x$ and $\kappa_{1z}$ are
shown on a  logarithmic scale.
}
\label{dispvsbeta}
\end{figure} 
%
We now discuss how the tilt of the Weyl cones directly influences
the mode diagrams and corresponding EM field characteristics. 
In Fig.~\ref{dispvsbeta}(a), the dimensionless propagation constant 
$\kappa_x$ is shown as a function of tilt $\beta$. 
For clarity of discussion later,
the
numbers label relevant parts of the dispersion diagram.
In  Fig.~\ref{dispvsbeta}(b) 
the wavevector in the WS, $\kappa_{1z}$ is
also shown over a wide range of $\beta$.
In contrast to what was observed in Fig.~\ref{dispvsomega}(b) when varying the frequency, 
here the wavevector in the WS monotonically increases along
the dispersion path $1\rightarrow 5$.
In Fig.~\ref{dispvsbeta}(c) the EM response
of both the diagonal (main plot) and off-diagonal (inset) 
components of $\rttensor{\epsilon}$
is shown 
over the same range of $\beta$. 
For the given  frequency, $\widetilde{\omega}=0.117$, 
both components $\epsilon_{\parallel}$
and  $\epsilon_{zz}$ are negative, with $\epsilon_{\parallel}\approx 0$.
Examining Fig.~\ref{dispvsbeta}(a), we see  that the main dispersion curve has the lower branch 
with points  $1\rightarrow 3$
close to the light line.
The upper branch from $3\rightarrow 5$ 
 is associated with substantially  larger 
propagation constants that increase rapidly with tilt.
The corresponding confinement factor
is  presented  in Fig.~\ref{dispvsbeta}(c).
As expected, the fraction of energy $\eta$ contained in the WS is
very small for points $1$ and $2$,  where
$\kappa_x\rightarrow 1$, 
reflecting the predominance of the energy density outside of the guiding semimetal layer.
There is a significant fraction of the energy density in the WS 
for modes occupying the upper dispersion curve
along the path $3\rightarrow 5$. Moreover,
 $\eta\rightarrow 1$
for conical tilts
in the vicinity of points 3 and 5, 
where $\partial \beta/\partial \kappa_x\rightarrow 0$  in Fig.~\ref{dispvsbeta}(a).
Thus, we see that the tilt of the Weyl cones
plays an
important role in determining the 
propagation and spatial localization properties 
for guided wave modes.

\begin{figure}[ht]
\centering
\includegraphics[width=\textwidth]{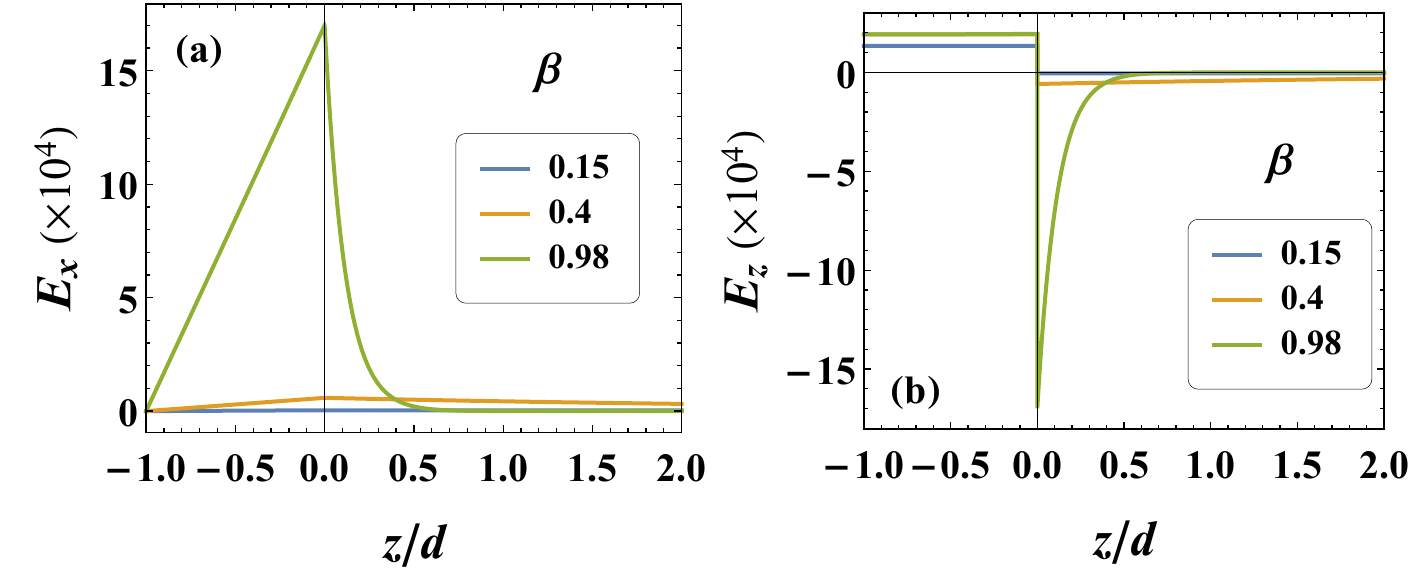}
\caption{
Panels (a) and (b)
display the electric field profiles as a function of
the normalized position $z/d$.
The legends depict the
range of conical tilts considered, which 
correspond to the upper branch of the mode diagram
in Fig.~\ref{dispvsbeta}(a) (with the labels $3\rightarrow 5$).
The frequency is identical to that used in Fig.~\ref{dispvsbeta}.
}
\label{fields1}
\end{figure} 
To study the guided modes in 
the ENZ regime further, we
keep the  wave frequency
the same as in Fig.~\ref{dispvsbeta}, 
corresponding to 
where $\epsilon_\parallel \approx 0$,
and investigate
the spatial behavior of the fields.
Thus, in
Fig.~\ref{fields1}, 
the components of the electric field
 $E_x$ and  $E_z$ are shown
 as a function of the scaled coordinate $z/d$
for varying  tilt of the Weyl cones.
The three values of the tilt parameter shown in the legend corresponds
to the points $3\rightarrow5$ of the mode diagram
in Fig.~\ref{dispvsbeta}(a).
Notably,
there is a linear variation in $E_x$, 
while $E_z$ is constant throughout the WS layer. 
The  ${\bm E}$-field for both components then
 decays once entering the vacuum region, as is required 
 for guided waves.
 For both components,  increasing the tilt
 results in a field enhancement, as
$\epsilon_{zz}$ becomes increasingly  more negative,
resulting in
modes
 with larger  $\kappa_x$ [see upper branch of Fig.~\ref{dispvsbeta}(a)].
To understand  this overall behavior explicitly, we
write the general expressions for the EM 
fields and then
take the limit $\epsilon_{\parallel} \rightarrow 0$, 
arriving  at:
\begin{align}
E_{x1} &\approx i Z_0 \sqrt{\kappa_x^2-1} \left(1+\frac{z}{d}\right), \label{eex1} \\
E_{z1} &\approx \frac{Z_0 \kappa_x \sqrt{\kappa_x^2-1}}{k_0 d(\kappa_x^2-\epsilon_{zz})},\label{eez1} \\
H_{y1}&\approx \frac{\epsilon_{zz}  \sqrt{\kappa_x^2-1}}{k_0 d(\epsilon_{zz}   -  \kappa_x^2)}, \label{hy1}
\end{align}
where the  factor $e^{i k_x x}$ has been suppressed.
The linear dependence seen in Fig.~\ref{fields1} for $E_{x1}$
is clearly evident from  Eq.~(\ref{eex1}),
while the spatial dependence in $E_{z1}$ and $H_{y1}$ drops out,
leading to the uniform component of the  electric field normal to the interface, 
seen in Fig.~\ref{fields1}(b).
Interestingly,  despite the fact that 
$E_{z1}$ is a function of $\kappa_x$ and $\epsilon_{zz}$, 
both of which 
depend  strongly on $\beta$ [see Fig.~\ref{dispvsbeta}(a) and \ref{dispvsbeta}(c)],
they tend to counteract one another, leading to the 
 weak $\beta$ dependence observed for $E_{z1}$ in Fig.~\ref{fields1}(b).
It is instructive to also examine the energy flow via the Poynting vector.
Within the ENZ regime,
we find,
\begin{align}
S_{x1} \approx \frac{Z_0}{2} \frac{\epsilon_{zz} \kappa_x (\kappa_x^2-1)}{ (k_0 d)^2 (\epsilon_{zz}-\kappa_x^2)^2},
\end{align}
which is independent of position in the WS.
The spatially averaged energy density in the WS is,
\begin{align}
\langle U_1 \rangle \approx   \mu (\kappa_x^2-1) 
\frac{ \omega\partial_\omega (\epsilon_{\parallel})   (k_0d)^2 (\epsilon_{zz} - \kappa_x^2)^2 
+ 3 (\epsilon_{zz}^2 + \kappa_x^2\partial_\omega (\omega\epsilon_{zz}))}
{12 (k_0d)^2 (\epsilon_{zz} - \kappa_x^2)^2}.
\end{align}
To obtain the speed of EM energy transport, 
it is then a simple matter to take the corresponding ratio via Eq.~(\ref{vtavg}).
The dispersion diagram can also be found analytically in the ENZ regime.
When $\epsilon_{\parallel}\approx 0$, the propagation constant
can be written:
\begin{align}
\kappa_x\approx\frac{\sqrt{\epsilon_{zz}[\epsilon_{zz}+2 (k_0 d)^2\pm \sqrt{\epsilon^2_{zz}+(2 k_0 d)^2(\epsilon_{zz}-1)}}]}{\sqrt{2} k_0 d},
\end{align}
which is a convenient expression that negates the need to numerically solve a transcendental equation for the
dispersion diagram.
%

\section{Conclusions}\label{conclusion}
We have analyzed the dispersion diagrams and electromagnetic field behavior of
a subwavelength  planar Weyl semimetal waveguide system where the Dirac cones in the touching points of valance and conduction bands possess the possibility to be tilted. We found that the electromagnetic response
for each diagonal component of the permittivity tensor  can be made
is vanishingly small,  creating an anisotropic epsilon-near-zero configuration. 
The admitted waveguide modes were presented in the form of
dispersion diagrams, where 
regimes of high energy-density confinement in the Weyl semimetal layer were found.  
We also showed how the net system power 
 can be manipulated  to vanish at critical points of the dispersion curves,
where the propagation velocity of electromagnetic energy is zero. Our results offer waveguides made of Weyl semimetals with controllable energy flow velocity and direction by materials and geometrical parameters. 

\section*{Funding}
K.H. is supported in part by ONR and a grant of HPC resources from the DOD HPCMP.  
M.A. is supported by Iran's National Elites Foundation (INEF).

\section*{Disclosures}
The authors declare that there are no conflicts of interest related to this article.

\appendix

\section{Calculation of $\gamma$: the off-diagonal component of the permittivity tensor $\rttensor{\epsilon}$}\label{apnxA}
When calculating the off-diagonal component $\gamma$ for arbitrary parameters,
it is convenient to separate Fermi surface and vacuum 
contributions to $\gamma(\omega)=\sum_{s=\pm}[\gamma^{(s)}_{FS}(\omega)+\gamma^{(s)}_0(\omega)]$, where
\begin{align}\label{dispgam}
\gamma^{(s)}_{FS} (\omega)&=
\frac{s \alpha}{\omega^2}\int_{0}^{\infty}\frac{p_{\perp}dp_{\perp}}{2\pi} \int_{-\Gamma-sv_FQ}^{\Gamma-sv_FQ}dp_z\frac{p_z}{p}\frac{i\omega_k}{p^2+\omega_k^2/4}[\Theta(\mu-\zeta_{s,+})-\Theta(\mu-\zeta_{s,-})-1],
\end{align}
and
\begin{align} \label{dispgam2}
\gamma^{(s)}_{0} (\omega)=\frac{s \alpha}{\omega^2}\int_{0}^{\infty}\frac{p_{\perp}dp_{\perp}}{2\pi}\int_{-\Gamma_{0}-sv_FQ}^{\Gamma_{0}-sv_FQ}dp_z\frac{p_z}{p}\frac{i\omega_k}{p^2+\omega_k^2/4},
\end{align}
where the following limit is taken in the integrals: $\omega_k\rightarrow\omega+i\delta$.
This ensures proper convergence when numerically evaluating Eqs.~(\ref{dispgam})-(\ref{dispgam2}).
We have also defined $\zeta_{s,t}\equiv tp+p_z\beta_s$,  
$p=\sqrt{p_z^2+p_{\perp}^2}$, and a momentum
cutoff along the $z$ axis, $\Gamma$. Generally, the cut-off $\Gamma$ is a function of the tilt parameter. 
Nevertheless, in our calculations, we choose a large enough cut-off and neglect the contribution of $\beta$ to $\Gamma$. 
The cut-off $\Gamma_0>v_F|Q|$ is introduced for the
correct definition of the vacuum contribution.

\section{Details of the coefficients}\label{appdxB}
Without loss of generality, we choose our EM field normalizations so that $r_1$=1. 
After matching the tangential electric and magnetic fields at the 
air/WS interface, and using the PEC boundary conditions, some straightforward algebra gives
the $r_2$ coefficient:
\label{coeff}
\begin{align}
r_2 = \frac{\kappa_x \gamma [\epsilon_{zz} \kappa_{0z} \kappa_- \kappa_+ (\cos q_- -\cos q_+) 
+ ( \kappa_x^2-\epsilon_{zz}) ( \kappa_+ \sin q_- - \kappa_- \sin q_+)]}
{
  \epsilon_{zz} [\kappa_+( \kappa_x^2 + \kappa_+^2 -\epsilon_\parallel ) 
  (\kappa_-   \cos q_- +  \kappa_{0z}   \sin q_-) - 
     \kappa_- ( \kappa_x^2 + \kappa_-^2-\epsilon_\parallel ) (\kappa_+ \cos q_+ + 
        \kappa_{0z} \sin q_+)]},
\end{align}
        and,
\begin{align}
a_1 &= Z_0 \frac{
  r_2 \epsilon_{zz} (\epsilon_\parallel - \kappa_x^2 - \kappa_-^2) (i \kappa_{0z} + \kappa_+) 
   + \kappa_x \gamma (i ( \kappa_x^2-\epsilon_{zz} ) + \epsilon_{zz} \kappa_{0z} \kappa_+)}
  {2 \epsilon_{zz} \kappa_x \kappa_+ ( \kappa_+^2  -\kappa_-^2)},\\
b_1 &= Z_0 \frac{
  r_2 \epsilon_{zz} (\epsilon_\parallel - \kappa_x^2 - \kappa_-^2) (  \kappa_+ -i \kappa_{0z}) 
   + \kappa_x \gamma (i ( \epsilon_{zz}-\kappa_x^2 ) + \epsilon_{zz} \kappa_{0z} \kappa_+)}
  {2 \epsilon_{zz} \kappa_x \kappa_+ ( \kappa_+^2  -\kappa_-^2)},\\
 c_1 &= Z_0 \frac{
  r_2 \epsilon_{zz} (\epsilon_\parallel - \kappa_x^2 - \kappa_+^2) (i \kappa_{0z} + \kappa_-) 
   + \kappa_x \gamma (i ( \kappa_x^2-\epsilon_{zz} ) + \epsilon_{zz} \kappa_{0z} \kappa_-)}
  {2 \epsilon_{zz} \kappa_x \kappa_- (\kappa_-^2- \kappa_+^2  )},\\
 d_1 &= Z_0 \frac{
  r_2 \epsilon_{zz} (\epsilon_\parallel - \kappa_x^2 - \kappa_+^2) (  \kappa_- -i \kappa_{0z}) 
   + \kappa_x \gamma (i ( \epsilon_{zz} -\kappa_x^2) + \epsilon_{zz} \kappa_{0z} \kappa_-)}
  {2 \epsilon_{zz} \kappa_x \kappa_- ( \kappa_-^2-\kappa_+^2 )}.
\end{align}
Once the dispersion equation [Eq.~(\ref{bigD})] is solved for the propagation constant $\kappa_x$,
the coefficients above can be determined to construct 
the EM fields.

\end{document}